\documentclass[conference]{IEEEtran}
\IEEEoverridecommandlockouts
\usepackage{cite}
\usepackage{amsmath,amssymb,amsfonts}
\usepackage{algorithmic}
\usepackage{graphicx}
\usepackage{textcomp}
\usepackage{xcolor}

\usepackage{color}
\usepackage{leftidx}
\usepackage{cite}
\usepackage{microtype}
\usepackage{cleveref}
\usepackage{graphicx,amssymb,amstext,amsmath}
\usepackage[ruled,vlined,lined,ruled,linesnumbered]{algorithm2e}
\usepackage{float}
\usepackage{bm}
\usepackage[font=small,skip=0pt]{caption}

\def\BibTeX{{\rm B\kern-.05em{\sc i\kern-.025em b}\kern-.08em
    T\kern-.1667em\lower.7ex\hbox{E}\kern-.125emX}}
\begin{document}
\title{Collaborative Computation Offloading in Wireless Powered Mobile-Edge Computing Systems }
\author{Binqi~He,~Suzhi~Bi,~Hong~Xing, and Xiaohui~Lin \\
College of Electronic and Information Engineering, Shenzhen University, Shenzhen, China 518060 \\
 E-mail:~hebinqi2017@email.szu.edu.cn, $\left\{\rm{bsz, hong.xing,xhlin} \right\}$@szu.edu.cn
 \vspace{-2ex}}

\maketitle

\begin{abstract}
This paper studies a novel user cooperation model in a wireless powered mobile edge computing system where two wireless users harvest wireless power transferred by one energy node and can offload part of their computation tasks to an edge server (ES) for remote execution. In particular, we consider that the direct communication link between one user to the ES is blocked, such that the other user acts as a relay to forward its offloading data to the server. Meanwhile, instead of forwarding all the received task data, we also allow the helping user to compute part of the received task locally to reduce the potentially high energy and time cost on task offloading to the ES. Our aim is to maximize the amount of data that can be processed within a given time frame of the two users by jointly optimizing the amount of task data computed at each device (users and ES), the system time allocation, the transmit power and CPU frequency of the users. We propose an efficient method to find the optimal solution and show that the proposed user cooperation can effectively enhance the computation performance of the system compared to other representative benchmark methods under different scenarios.
\end{abstract}

\IEEEpeerreviewmaketitle

\section{Introduction}
The explosive growth of Internet of Things (IoT) and 5G communication technologies have driven the increasing computing demands for wireless devices. In the meantime, an IoT device (e.g., sensor) often carries a capacity-limited battery and an energy-saving low-performance processor under the consideration of the stringent device size constraint and production cost. Recently, wireless powered mobile-edge computing (MEC) has emerged as a promising technique to solve the above problems \cite{2017:Mao,2017:Wu,2019:XingH,2018:MY}. In a wireless powered MEC system, WDs are powered by means of wireless power transfer (WPT) \cite{2016:Bi} and can offload intensive computations to the edge servers located at the radio access networks \cite{2019:ZLiang,2019:WuW}. Such an integration of MEC and WPT technologies effectively solves the limitations of both on-chip energy and computing capability of IoT devices.

Existing studies on wireless powered MEC mostly concern the joint offloading and resource allocation to enhance the computing performance of the system. For instance, \cite{2016:CK} considered a single user following a binary offloading policy that the computation task is non-partitionable and must be offloaded as a whole. It then finds the optimal offloading strategy by optimizing the computation and communication resource to maximize the probability of successful computation. Such design was extended to multiuser MEC systems in \cite{2018:Bi}, which considered a binary computation offloading policy and jointly optimized the individual computing mode selection and system transmission time allocation to maximize the processed task data in a given time frame. On the other hand, \cite{2018:Xu} considered a partial offloading methods where the computation task of each user can be arbitrarily partitioned and executed separately at both the local device and remote edge server. It then jointly optimizes the WPT transfer, spectral and computing resource allocation to minimize the server energy consumption under user computing delay constraint. Several other works have also considered the designs of wireless powered MEC system under random task arrivals \cite{2017:Xu}, using machines learning based structure \cite{2019:Huang}, and for UAV-enabled scenarios \cite{2018:Hua,2018:Zhou}.

An inherent problem in wireless powered systems is severe user near-far unfairness caused by drastic attenuation of the RF energy signal over distance. User cooperation is an effective solution to enhance the overall system communication and computing performance. For instance, \cite{2014:Ju2,2017:YLN,2017:ZMQ} proposed various user cooperation methods in wireless powered communication networks. Under the wireless powered MEC paradigm, \cite{2018:Hu} considered a two-user scenario where one acts as the relay to forward the other's computation offloading to the AP. It then minimizes the total transmit energy of the AP by jointly optimizing the power and time allocation under computation performance constraint. \cite{2018:DF} investigated a computation cooperation method, in which an active-computing user can offload part of its task to multiple helpers for remote execution. Furthermore, \cite{2018:Cao} considered exploiting joint computation and communication cooperation where a helper node acts not only as a communication relay to the AP but also a computing agent that can compute the user's task locally. Unlike \cite{2018:Cao} that considers a dedicated helper, in this paper, we consider the case that the relay helper also has its own task to process, thus needs to carefully allocate its resource, e.g., computing power and energy, to both help the other user and compute its own task. Such situation requires joint optimization of the task offloading of the two users, user computing resource allocation, and the system-level wireless resource allocation to maximize the overall system computing performance.

\begin{figure}
  \centering
   \begin{center}
      \includegraphics[width=0.45\textwidth]{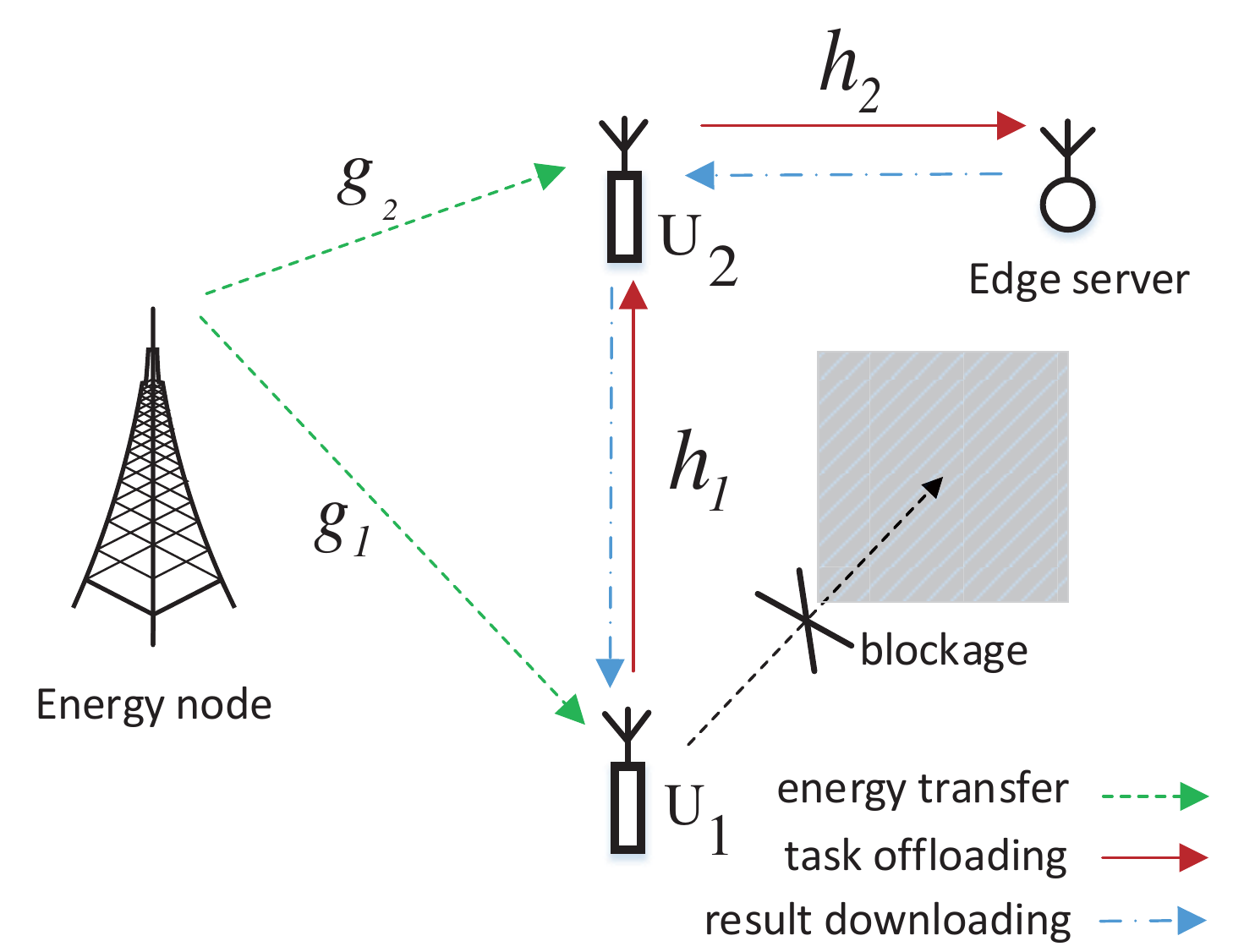}
   \end{center}
  \caption{The schematic of the considered cooperation in a wireless powered MEC system. }
  \label{Fig.1}
\end{figure}

In this paper, we investigate a novel two-user cooperation method in wireless powered MEC system. As shown in Fig.~\ref{Fig.1}, we consider users powered by an energy node by means of WPT. Using the harvested energy, the two users then compute their own tasks assisted by an edge server (ES), i.e., both users' tasks can be offloaded for edge execution. In particular, we consider that one user ($\rm{U_1}$) is blocked from direct communication to the ES, such that the other user ($\rm{U_2}$) can relay $\rm{U_1}$'s task to the MEC server. Meanwhile, we allow $\rm{U_2}$ to allocate part of its resource to compute $\rm{U_1}$'s task to reduce the offloading communication delay to the edge server. The main contributions of this paper are summarized as follows:
\begin{itemize}
  \item We present a new user collaboration model in wireless powered MEC, where the helping user $\rm{U_2}$ acts both as the communication relay and the computing agent for the other user $\rm{U_1}$. In particular, we consider a partial offloading scheme, such that the task data of $\rm{U_1}$ can be partitioned at three parts and computed at itself, $\rm{U_2}$, and the ES, respectively. The task of $\rm{U_2}$, on the other hand, can be partitioned and computed both locally and offloaded to the ES, respectively.
  \item We formulate an optimization problem to maximize the weighted sum-computation-rates (WSCR) of the two users, which is an direct measure of the data processing capability of the system. This involves a joint optimization of task partitions, user resource allocation (CPU frequency and transmit power), system-level time allocation (on WPT and data transmission). We propose an efficient method to solve the problem optimally.
  \item We compare the performance of the proposed method with two representative benchmark methods, where either computation or communication cooperation is absent. We show that the proposed method significantly outperforms the two benchmarks especially when task offloading is costly for either user, e.g., weak inter-user or user-ES channel. We also show that not only $\rm{U_1}$, but also the helper $\rm{U_2}$ can benefit from the joint communication and computation collaborations.
\end{itemize}
%The notations of channel gains are given in Fig.~\ref{Fig.1}. For simplicity of illustration, we assume that all the channels are reciprocal and under quasi-static flat-fading, where the channel gains remain constant during each time frame but vary from one frame to another.
\section{System Model}
\subsection{Protocol Description and Channel Model}
As shown in Fig.~\ref{Fig.1}, we consider a wireless powered MEC system consisting of an energy node (EN), two energy-harvesting users and an edge server (ES), where each device is equipped with a single antenna. In particular, we assume that the EN has constant power supply and transfers wireless power to $\rm{U_1}$ and $\rm{U_2}$. The two users have no other embedded energy source and rely on the harvested energy to compute their own tasks. In particular, both users can offload part of their tasks to the ES following a partial offloading policy. Due to the hardware constraint, each user reuses its antenna for both energy harvesting and task data transmission in an time division duplexing (TDD) manner \cite{2013:Zhang}. Specifically, we assume that the direct communication between $\rm{U_1}$ and ES is blocked, such that $\rm{U_2}$ may serve as the relay to forward the computation offloading of $\rm{U_1}$. Besides, we also assume that the relay user $\rm{U_2}$ may allocate part of its computation resource to help compute $\rm{U_1}$'s task, for instance, when task offloading to the server is costly under deep channel fading. After the ES finish computing the received task(s), it sends the results back to $\rm{U_2}$, and so does $\rm{U_2}$ to $\rm{U_1}$ subsequently. For simplicity, all the channels are assumed to be independent and reciprocal and follow quasi-static flat-fading, such that all the channel coefficients remain constant during each block transmission time. As illustrated in Fig.~\ref{Fig.1}, we use ${g_i}$, ${h_i},~i = 1, 2$, to denote the corresponding channel gains.

Notice that in the above cooperative communication and computation model, $\rm{U_1}$'s task can be executed at $\rm{U_1}$, $\rm{U_2}$ and ES, while $\rm{U_2}$'s task can be executed at $\rm{U_2}$ and ES.

\subsection{System Time Allocation}

\begin{figure}
  \centering
   \begin{center}
      \includegraphics[width=0.45\textwidth]{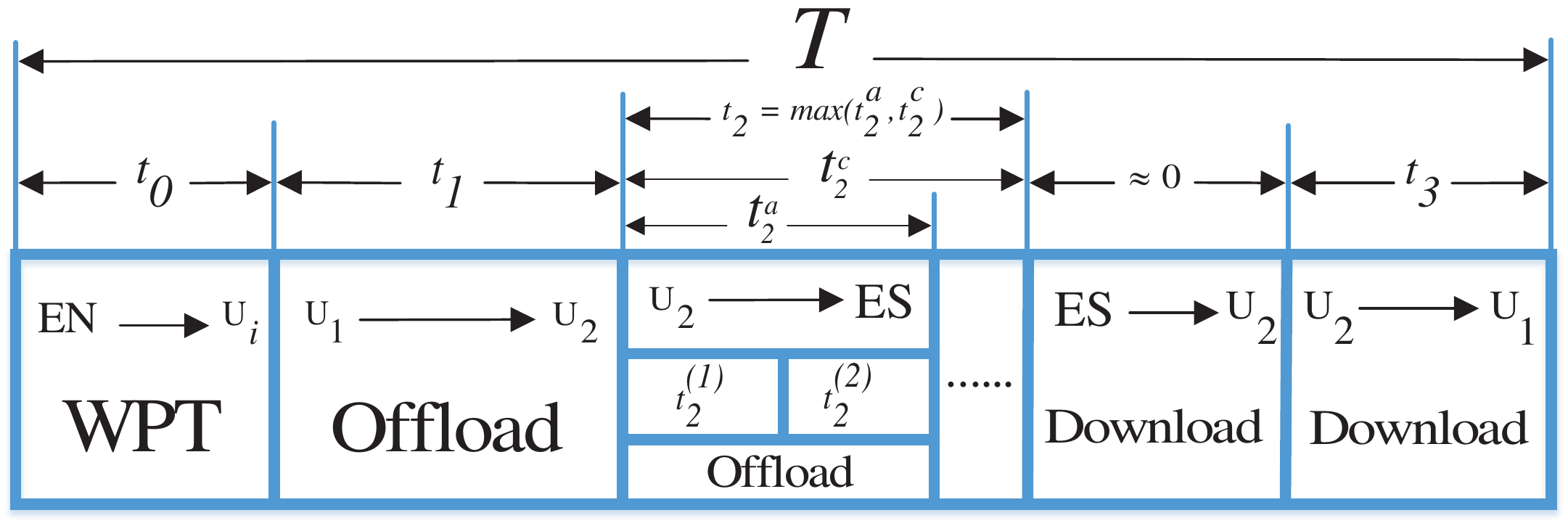}
   \end{center}
  \caption{Time allocation in the wireless powered MEC network.}
  \label{Fig.2}
\end{figure}

For a tagged time frame of duration $T$, we show the detailed transmission time allocation in Fig.~\ref{Fig.2}. At the beginning of a time frame, the EN transfers wireless power to $\rm{U_1}$ and $\rm{U_2}$ for a duration of ${{t_0}}$ with a fixed transmit power ${p_0}$. After the WPT period, $\rm{U_1}$ offloads part of its computation task to $\rm{U_2}$ in the subsequent period ${t_1}$. In the next time slot of duration ${{t_2}}$, $\rm{U_2}$ first relays the received $\rm{U_1}$'s task with power $p_2^{(1)}$ for $t_2^{(1)}$ amount of time to the ES and then offloads its own task to the server with power $p_2^{(2)}$ for $t_2^{(2)}$ amount of time. We denote $t_2^a = t_2^{(1)} + t_2^{(2)}$ as the total task offloading time of $\rm{U_2}$. After receiving the tasks from $\rm{U_2}$, the server computes and returns the results to $\rm{U_2}$. Here, we assume the ES has much stronger transmit power and computing capability than the energy-harvesting users, thus neglect the time consumed on computing at the edge server and data transmission back to $\rm{U_2}$ (similar to the assumptions in \cite{2018:Xu,2018:Bi}). Notice that $\rm{U_2}$ can also compute part of $\rm{U_1}$'s task locally. Here we assume that $\rm{U_2}$ can only compute one task at a time, such that it computes $\rm{U_1}$'s task immediately after receiving it. The total computation time of $\rm{U_2}$ spent on computing $\rm{U_1}$'s task is denoted as $t_2^c$. We use $t_2$ to denote the duration from $\rm{U_2}$ starts to receive the task offloading from $\rm{U_1}$ until it starts to send the results back to $\rm{U_1}$. Evidently, $t_2$ equals the longer duration between $t_2^c$ and $t_2^a$, i.e.,
\begin{equation}\label{1}
  {t_2} = \max(t_2^c,t_2^a).
\end{equation}

After $\rm{U_2}$ receiving the computation result of $\rm{U_1}$ from the server, it first combines the server's feedback with the locally computed result for $\rm{U_1}$, and then sends back to $\rm{U_1}$. We denote the corresponding transmission time and transmit power as $t_3$ and $p_3$, respectively. Overall, we have a total system time allocation constraint
\begin{equation}\label{2}
\left\{ \begin{array}{l}
{t_0} + {t_1} + {t_2} + {t_{\rm{3}}} = T,\\
{t_0},{t_1},{t_2},t_2^{(1)},t_2^{(2)},t_2^c,{t_{\rm{3}}} \ge 0.
\end{array} \right.
\end{equation}
In the following section, we derive the computation performance of both users and formulate the problem to maximize the system's data processing capability.

\section{Computation Performance Analysis and Problem Formulation}

\subsection{Wireless Energy Transmission}
In the first part of a tagged time frame, the EN broadcasts wireless energy to each user for a duration of ${{t_0}}$. We let ${{p_0}}$ denote the fixed transmit power of the EN such that the amount of energy harvested by $\rm{U_i}$ can be expressed as
 \begin{equation}\label{3}
   {Q_i} = \mu {g_i}{t_0}{p_0}, ~i=1,2,
 \end{equation}
where $0 < \mu  < 1$ denotes the energy harvesting efficiency assumed equal for both users.\footnote{Although a single energy harvesting circuit exhibits non-linear energy harvesting property due to the saturation effect of circuit, it is shown that the non-linear effect can be effectively rectified by using multiple energy harvesting circuits concatenated in parallel, resulting in a sufficiently large linear conversion region in practice \cite{2019:JK}.}

With the harvested energy, the two users compute their tasks locally or/and offload the computations to other devices for remote execution. We let ${b_{ij}}$ denote the amount of the $i$-th user's task data processed at the $j$-th device within a tagged time block (in bits), where the ES is indexed as 0. Accordingly, the total number of processed bits of the two users are denoted as ${b_1} = {b_{11}} + {b_{12}} + {b_{10}}$ and ${b_2} = {b_{22}} + {b_{20}}$, respectively. The major performance metric is computation rates of the two users (in bits/second), which are expressed as ${x_1} = \frac{{{b_1}}}{T}$ and ${x_2} = \frac{{{b_2}}}{T}$, respectively. Without loss of generality, we assume $T=1$ in the following analysis, such that we use $\mathbf{b}$ and $\mathbf{x}$ interchangeably. In the following, we derive the expressions of $b_1$ and $b_2$.

\subsection{Individual Local Computation}
Recall that the $i$-th user computes $b_{ii}$ amount of its own task locally, where $i = 1,2.$ Let $\phi > 0$ denotes the number of CPU cycles required for computing one input bit, which is assumed equal for both users without loss of generality. Besides, we denote ${f_i} \in [0,f_i^{max}]$ as the $i$-th user's CPU computing speed (cycles per second), where $f_i^{max }$ is the maximum CPU frequency. Notice that each user can compute throughout the time block without being interrupted by information or energy transmissions. We denote $t_{ii}$ as the corresponding local computation time of the $i$-th user.
Then, the local computation rate of $\rm{U_i}$ is
\begin{equation}\label{4}
  {b_{ii}} = \frac{{{f_i}{t_{ii}}}}{{\phi}},~i = 1, 2.
\end{equation}
Meanwhile, the corresponding energy consumption is \cite{2018:Bi}
\begin{equation}\label{5}
  E_{i,i}^{loc} = {k_i}{f_i}^3{t_{ii}},~i = 1, 2,
\end{equation}
where $k_i$ is the effective capacitance coefficient that depends on the chip architecture at user $i$.
\subsection{Computation Offloading to Edge Server}
Besides local computation, each user can also offload part of their computing task to the edge for remote execution. The computation rate due to edge computation of the two users are denoted as $b_{10}$ and $b_{20}$. In particular, the amount of task data that is offloaded from $\rm{U_1}$ to $\rm{U_2}$ is constrained by
\begin{equation}\label{6}
    {b_{12}} + {b_{10}} \le {t_1}B{\log _2}\left( {1 + \frac{{{h_1}{p_1}}}{{\Gamma {N_0}}}} \right),
\end{equation}
where $B$ denotes the system bandwidth, $N_0$ represents the power of receiver noise, and $\Gamma  \ge 1$ is a constant term accounting for the gap from the channel capacity due to a practical modulation and coding scheme. Meanwhile, the energy consumption on transmitting the information of $\rm{U_1}$ is
\begin{equation}\label{7}
    E_1^{off} = {t_1}{p_1}.
\end{equation}
After receiving $b_{10}$ from $\rm{U_1}$, $\rm{U_2}$ acts as a relay to help $\rm{U_1}$ offload ${b_{10}}$ task data to the ES with the time duration $t_2^{(1)}$ and transmit power $p_2^{(1)}$. Therefore, ${b_{10}}$ is constrained by
\begin{equation}\label{8}
    {b_{10}} \le t_2^{(1)}B{\log _2}\left( {1 + \frac{{{h_2}p_2^{(1)}}}{{\Gamma {N_0}}}} \right).
\end{equation}
Then $\rm{U_2}$ offloads its own ${b_{20}}$ task data to the ES with the time duration $t_2^{(2)}$ and power $p_2^{(2)}$, such that we have
\begin{equation}\label{9}
    {b_{20}} \le t_2^{(2)}B{\log _2}\left( {1 + \frac{{{h_2}p_2^{(2)}}}{{\Gamma {N_0}}}} \right).
\end{equation}

The energy consumption of $\rm{U_2}$ on offloading the tasks to the edge server is
\begin{equation}\label{10}
    E_2^{off} = t_2^{(1)}p_2^{(1)} + t_2^{(2)}p_2^{(2)}.
\end{equation}

\subsection{Collaborative Computing at $U_2$}
Besides forwarding the received computation task from $\rm{U_1}$ to the edge server, $\rm{U_2}$ also executes part of $\rm{U_1}$'s task, whose amount is denoted as $b_{12}$. We denote $t_2^c$ and $f_{2c}$ as the corresponding computation time and CPU frequency for processing $b_{12}$, which are related as
\begin{equation}\label{10a}
  {b_{12}} = \frac{{{f_{2c}}t_2^c}}{\phi }.
\end{equation}
Besides, the corresponding consumed energy by $\rm{U_2}$ is
\begin{equation}\label{11}
  E_{2,1}^{loc} = {k_2}{({f_{2c}})^3}t_2^c.
\end{equation}
Notice that the total local computation time of $\rm{U_2}$ is constrained by
\begin{equation}\label{12}
  t^c_{2} + t_{22} \leq 1.
\end{equation}
In time slot $t_3$, $\rm{U_2}$ first combines the computation outcomes of $\rm{U_1}$ received from the edge server and produced locally, and then transmits the result back to $\rm{U_1}$. Here, we assume the amount of computation outcome is proportional to the task input, such that $\rm{U_2}$ needs to send $\nu(b_{10} + b_{12})$ bits back to $\rm{U_1}$, where $\nu  < 1$ is a fixed parameter. We denote the transmit power used for feeding back the result to $\rm{U_1}$ as $p_3$. Evidently, the transmission rate cannot exceed the channel capacity, i.e.,
\begin{equation}\label{13}
  \nu ({b_{10}} + {b_{12}}) \le {t_3}B{\log _2}\left( {1 + \frac{{{h_1}{p_3}}}{{\Gamma {N_0}}}} \right).
\end{equation}
Besides, the consumed energy is denoted as ${E_{21}} = {t_3}{p_3}$. From the above discussion, the consumed energy of the two users are constrained by the total individual harvested energy
\begin{equation}\label{14}
  E_{1,1}^{loc} + E_1^{off} \le {Q_1},
\end{equation}
\begin{equation}\label{15}
  E_{2,2}^{loc} + E_{2,1}^{loc} + E_2^{off} + {E_{21}} \le {Q_2}.
\end{equation}

\subsection{Problem Formulation}
In this paper, we are interested in maximizing the weighted sum-computation-rates (WSCR) of the two users by jointly optimizing the number of task bits processed at each device ($\mathbf{b}$), user resource allocation (CPU frequency $\bm{f} = \left\{{f_1,f_2,f_{2c}}\right\}$ and transmit power $\bm{p} = \{ {p_1},p_2^{(1)},p_2^{(2)},{p_3}\} $), and system-level time allocation (on WPT and data transmission $\bm{t} = \{ {t_0},{t_1},t_2^{(1)},t_2^{(2)},t_2^c,{t_3}\} $. Thus, the problem can be mathematically expressed as
\begin{subequations}
   \begin{align*}
   (\bm{P1}): \ & \underset{\bm{t,p,f,b}}{\text{maximize}} & & {w_1}({b_{11}} + {b_{12}} +{b_{10}}) + {w_2}({b_{22}} + {b_{20}})   \\
    & \text{subject to} & & (\ref{1}),(\ref{2}),(\ref{4}),(\ref{6}),(\ref{8}),(\ref{9}),(\ref{10a}),(\ref{12})-(\ref{15})
   \end{align*}
\end{subequations}
%\begin{subequations}
%   \begin{align*}
%   (\bm{P1}): \ & \underset{\bm{t,p,f}}{\text{maximize}} & & {w_1}({X_{11}} + {X_{12}} +{X_{10}}) + {w_2}({X_{22}} + {X_{20}})   \tag{21a} \\
%    & \text{subject to} & & {t_0} + {t_1} + {t_2} + {t_{\rm{3}}} \le T, \tag{21b} \\
%    & & & {t_2} = \max \{ t_2^c,t_2^a\}, \tag{20c} \\
%    & & & {f_{2c}}t_2^c + t_2^{(1)}B{\log _2}(1 + \frac{{{{\left\| {{h_1}} \right\|}^2}p_2^{(1)}}}{{{N_0}}}) \\
%    & & & \le {t_1}B{\log _2}(1 + \frac{{{{\left\| {{h_1}} \right\|}^2}{p_1}}}{{{N_1}}}), \tag{20d} \\
%    & & & t_2^{(1)}p_2^{(1)} + t_2^{(2)}p_2^{(2)} + {k_2}f_2^3(T - {t_0} - t_2^c) \\
%    & & & +{k_{2c}}f_{2c}^3t_2^c + {t_3}{p_3} \le \mu {\left| {{g_2}} \right|^2}{t_0}{p_0}, \tag{20e} \\
%    & & & {t_1}{p_1} + {k_1}f_1^3(T - {t_0}) \le \mu {\left| {{g_1}} \right|^2}{t_0}{p_0}, \tag{20f} \\
%    & & & 0 \le {t_i} \le T,0 \le {f_i} \le f_i^{\max } \tag{20g}.
%   \end{align*}
%\end{subequations}
Here ${w_i} > 0$ denotes the weight associated with $\rm{b_i}$ and $w_1+w_2 =1$. Note that problem $\bm{P1}$ is a non-convex optimization problem in the above form, e.g., due to the multiplicative terms in (\ref{4}) and (\ref{13}), and the non-concave functions in (\ref{8}) and (\ref{9}). In the following, we provide the optimal solution to $\bm{(P1)}$.

\section{Optimal Solution to $\bm{(P1)}$}
In this section, we study the optimal solution to \bm{$P1$}, which is transformed into a convex problem and accordingly solved optimally with off-the-shelf convex algorithms, e.g., interior point method. To begin with, we introduce an auxiliary vector, $\bm{\tau}  = \left[ {{\tau _0},{\tau _1},\tau _2^{(1)},\tau _2^{(2)},{\tau _{\rm{3}}}} \right]$, where ${\tau _0} = {t_0}{p_0},{\tau _1} = {t_1}{p_1},\tau _2^{(1)} = t_2^{(1)}p_2^{(1)},\tau _2^{(2)} = t_2^{(2)}p_2^{(2)},$ and ${\tau _3} = {t_3}{p_3}$. Then, we have that
\begin{equation}\label{17}
\left\{ {\begin{array}{*{20}{l}}
{{p_0} = \frac{{{\tau _0}}}{{{t_0}}},{p_1} = \frac{{{\tau _1}}}{{{t_1}}},{p_3} = \frac{{{\tau _3}}}{{{t_3}}},}\\
{p_2^{(1)} = \frac{{\tau _2^{(1)}}}{{t_2^{(1)}}},p_2^{(2)} = \frac{{\tau _2^{(2)}}}{{t_2^{(2)}}}.}
\end{array}} \right.
\end{equation}
By substituting (\ref{17}) into (\ref{6}) and (\ref{13}), we have
\begin{equation}\label{18}
  {b_{10}} + {b_{12}} \le {t_1}B{\log _2}\left( {1 + {\rho _1}\frac{{{\tau _1}}}{{{t_1}}}} \right),
\end{equation}
\begin{equation}\label{19}
  \nu ({b_{10}} + {b_{12}}) \le {t_3}B{\log _2}\left( {1 + {\rho _1}\frac{{{\tau _3}}}{{{t_3}}}} \right),
\end{equation}
and into (\ref{8}), (\ref{9}), we have
\begin{equation}\label{20}
  {b_{10}} \le t_2^{(1)}B{\log _2}\left( {1 + {\rho _2}\frac{{\tau _2^{(1)}}}{{t_2^{(1)}}}} \right),
\end{equation}
\begin{equation}\label{21}
  {b_{20}} \le t_2^{(2)}B{\log _2}\left( {1 + {\rho _2}\frac{{\tau _2^{(2)}}}{{t_2^{(2)}}}} \right).
\end{equation}
Notice that the above constraint (\ref{18})-(\ref{21}) are jointly convex in \bm{$\left\{\tau, t, b\right\}$}. From \cite{2018:Bi}, each of the two energy-constrained users should compute throughout the time block to maximize the local computation rate. This indicates that $t_{11}=1$ and $t^c_2 + t_{22} =1$ holds. Accordingly, constraints (\ref{14}) and (\ref{15}) can be respectively transformed into
\begin{equation}\label{22}
  {\tau _1} + {k_1}f_1^3 \le {\rho _3}{\tau _0},
\end{equation}
\begin{equation}\label{23}
  \tau _2^{(1)} + \tau _2^{(2)} + {k_2}f_2^3(1 - t_2^c) + {k_{2c}}f_{2c}^3t_2^c + {\tau _3} \le {\rho _4}{\tau _0},
\end{equation}
where ${\rho _1} = \frac{{{h_1}}}{{\Gamma {N_0}}}$, ${\rho _2} = \frac{{{h_2}}}{{\Gamma {N_0}}}$, ${\rho _3} = \mu {g_1}$ and ${\rho _4} = \mu {g_2}$ are constant parameters.
Similarly, constraint (\ref{4}) reduces to
\begin{equation}\label{24}
  {b_{11}} = \frac{1}{\phi }{f_1}, ~ {b_{22}} = \frac{{{f_2}(1 - t_2^c)}}{\phi }.
\end{equation}
From the above discussion, problem $\bm{(P1)}$ is equivalently transformed into
\begin{subequations}
   \begin{align*}
   (\bm{P2}): \ & \underset{\bm{t,\tau,b,f}}{\text{maximize}} & & {w_1}({b_{11}} + {b_{12}} +{b_{10}}) + {w_2}({b_{22}} + {b_{20}})   \\
    & \text{subject to} & & (\ref{1}),(\ref{2}),(\ref{10a}),(\ref{18})-(\ref{24})
   \end{align*}
\end{subequations}
%\begin{equation}\label{27}
%    \begin{split}
%        (\bm{P2}): &\mathop {\max }\limits_{\bm{t,\tau,b,f}} \quad ({b_{11}^ *} + {b_{12}} +{b_{10}}){w_1} + ({b_{22}^ *} + {b_{20}}){w_2} \\
%        &\text{s. t.} \qquad (\ref{1}),(\ref{2}),(\ref{18})-(\ref{23})
%    \end{split}
%\end{equation}
Notice that the above problem is still non-convex because of the multiplicative terms in (\ref{23}) and (\ref{24}). However, it becomes a convex problem once we fix $t_2^c \in [0,1)$, thus can be efficiently solved given $t^c_2$. For simplicity, we denote the optimal value of \bm{$(P2)$} given $t^c_2 = z$ as $S(z)$, where $z\in[0,1]$. Therefore, \bm{$(P2)$} reduces to a one-dimensional search to find the optimal $t^c_2$ that maximizes $S(t^c_2)$, where we devise a golden-section search over the scalar variable $t^c_2$ in Algorithm.~\ref{alg1}. After solving \bm{$(P2)$} optimally,  we can retrieve the optimal transmit power to \bm{$(P1)$} from (\ref{17}).
%\begin{algorithm}
%\footnotesize
% \SetAlgoLined
% \SetKwData{Left}{left}\SetKwData{This}{this}\SetKwData{Up}{up}
%
% \SetKwFunction{Union}{Union}\SetKwFunction{Find Compress}{Find Compress}
% \SetKwInOut{Input}{input}\SetKwInOut{Output}{output}
% \Input{time duration $T=1$ and other system parameters.}
% \Output{the optimal time allocation of $\{{t_0},{t_1},{t_2^{(1)}},{t_2^{(2)}},{t_2^c},{t_3}\}$}
% Initialize: $\bm{S}\leftarrow{objective function}$, {$\Delta  \leftarrow $ small positive step size;}\\
% Set [a,b] = [0,0.9], $t_2^c\leftarrow{a}$;
%
%    \While{$t_2^c \le b$}
%    {Calculate and compare $\bm{S}\left| {_{t_2^c = a}} \right.$ and $\bm{S}\left| {_{t_2^c = a + \Delta }} \right.$ \\
%        \Repeat{$\bm{S}(a + \Delta ) - \bm{S}(a)| \le \sigma $}{\If{$\bm{S}(a + \Delta ) > \bm{S}(a)$}
%                {$t_2^{c*} \leftarrow a + \Delta$  \\
%
%                \Else {$t_2^{c*} \leftarrow a$}
%                                                    }
%            $\bm{Update}$ a and b according to the Golden section search method;
%                                                    }}
%
%\textbf{Return} $\{ t_2^{c*}, {t_0}^*, {t_1}^*, t_2^{{(1)}^*}, t_2^{{(2)}^*}, {t_3}^*  \}$ .
%\caption{Optimal time allocation solution to $\bm{(P2)}$}
%\label{alg1}
%\end{algorithm}

\begin{algorithm}
\footnotesize
 \SetAlgoLined
 \SetKwData{Left}{left}\SetKwData{This}{this}\SetKwData{Up}{up}
 \SetKwRepeat{do While}{do}{while}
 \SetKwFunction{Union}{Union}\SetKwFunction{Find Compress}{Find Compress}
 \SetKwInOut{Input}{input}\SetKwInOut{Output}{output}
% \Input{objective function \bm{$S$} and other system parameters.}
% \Output{the optimal $t_2^{c*}$ and the system time allocation.}
 \bm{${\rm{Initialize}}$}: search interval [$a_0$,$a_1$]=[0,1], golden section factor $\sigma =0.618$, accuracy control $\varepsilon =10^{-4}$, number of iterations $m=1$
\;
Let ${\lambda _m} = {a_0} + (1 - \sigma )\cdot ({a_1} - {a_0}),{\gamma_m} = {a_0} + \sigma \cdot ({a_1} - {a_0})$, calculate $S(\lambda_m)$ and $S(\gamma_m)$ by solving a convex problem in (P2). \\
\textbf{while} $|{a_1} - {a_0}| > \varepsilon $ \textbf{do} \\
\quad \quad \textbf{if} $S(\lambda_m){<}S(\gamma_m)$ \textbf{then} \\
\quad \quad \quad \quad ${a_0}=\lambda_m,{a_1}={a_1},~\lambda_{m+1}=\gamma_m$; \\
\quad \quad \quad \quad $\gamma_{m+1}={a_0}+\sigma \cdot ({a_1}-{a_0});$ \\
\quad \quad \quad \quad \bm{${\rm{Update}}$} $S(\lambda_{m+1})\leftarrow S(\gamma_m)$ and calculate $S(\gamma_{m+1})$.  \\
\quad \quad \textbf{else}\\
\quad \quad \quad \quad ${a_0}={a_0},{a_1}=\gamma_m,~\gamma_{m+1}=\lambda_m$; \\
\quad \quad \quad \quad $\lambda_{m+1}={a_0}+(1-\sigma )\cdot ({a_1}-{a_0})$; \\
\quad \quad \quad \quad \bm{${\rm{Update}}$} $S(\gamma_{m+1})\leftarrow S(\lambda_m)$ and calculate $S(\lambda_{m+1})$. \\
\quad \quad \textbf{end if} \\
\quad \quad  $m=m+1$; \\
\textbf{end while}\\
\textbf{if} $S(\lambda_m){<}S(\gamma_m)$ \textbf{then}\\
\quad \quad $t_2^{c*}=\gamma_m$\\
\textbf{else}\\
\quad \quad $t_2^{c*}=\lambda_k$\\
\textbf{end if}\\
\textbf{Return} $t_2^{c*}$ and the corresponding optimal solutions to \bm{$(P2)$} given $t^{c*}_2$.
\caption{Optimal solution to $\bm{(P2)}$}
\label{alg1}
\end{algorithm}

\vspace{0.5cm}

\section{Simulation Results}
%\begin{figure}
%  \centering
%  \begin{center}
%    \includegraphics[width=0.48\textwidth]{Fig3.pdf}
%  \end{center}
%  \caption{The impact of ${t^c_2}$ to the optimal WSCR performance. }
%  \label{Fig.3}
%\end{figure}
In this section, we use simulations to evaluate the performance of the proposed cooperation method. In all simulations, we set ${p_0} = 3~W$, $\mu  = 0.7$, ${N_0} = {10^{ - 10}}~W$, $\Gamma = 1 $. For simplicity of illustration, the wireless channel gain between two devices, denoted by $h$, follows a path-loss model $h = {G_A}{(\frac{{3 \cdot {{10}^8}}}{{4\pi {d}f}})^\lambda }$, where $d$ denotes the distance of the two devices considered, $f=915~MHz$ denotes carrier frequency, $\lambda$ denotes the path loss factor, and ${G_A} = 2$ denotes the antenna power gain. Unless otherwise stated, we set $\lambda=2.5$, ${w_1}  = 0.7$, ${w_2}  = 0.3$, the EN-$\rm{U_1}$ and EN-$\rm{U_2}$ distances as ${d_{E,1}} = 6m$ and ${d_{E,2}} = 4m$, the $\rm{U_2}$-ES distance as ${d_{20}} = 10m$, and the $\rm{U_1}$-$\rm{U_2}$ distance as ${d_{12}} = 4m$. Besides, we set equal computing efficiency parameter ${k_i} = {10^{ - 26}}$, $f_i^{max} = 3\cdot 10^6$, and $\phi  = 100$ for both devices. For data offloading, we set the bandwidth $B = 10~kHz$ and $\nu  = 0.5$. For performance comparison, we consider the following representative benchmark methods:
\begin{enumerate}
  \item Communication cooperation only (Benchmark 1): $\rm{U_2}$ only acts as a relay to help $\rm{U_1}$ offload parts of its task to ES , i.e., $b_{12} = 0$.
  \item Computation cooperation only (Benchmark 2): $\rm{U_2}$ only acts as a computing agent to help $\rm{U_1}$ process parts of its task, i.e., $b_{10} = 0$.
\end{enumerate}

%We first show in Fig.~\ref{Fig.3} the impact of ${t^c_2}$ factor to the WSCR performance. Here, we set ${g_1} = 2.19 \times {10^{ - 5}},{g_2} = 2.19 \times {10^{ - 4}},{h_1} = 9.41 \times {10^{ - 5}},{h_2} = 9.41 \times {10^{ - 5}}$, and change the value of $t_2^c$ from $0.1$ to $0.9$. It can be seen that when $t_2^c \in (0.6,0.8)$, we can get relatively higher value of the WSCR, $t_2^c$ increases from 0 and reaches the maximum around 0.7. We know $t_2^c$ is the time of ${U_1}'s$ task computing in ${U_2}$, which is a parallel time with $t_2^a$. But as we further increase the value of $t_2^c$, the WSCR of system is not always better. That's because the constrain of (\ref{12}), once $t_2^c$ keeps increasing, ${t_{22}}$ and other time allocation would be reduced. So we assume a fixed $t_2^c=0.7$ in the following simulations.
\begin{figure}
  \centering
  \begin{center}
    \includegraphics[width=0.45\textwidth]{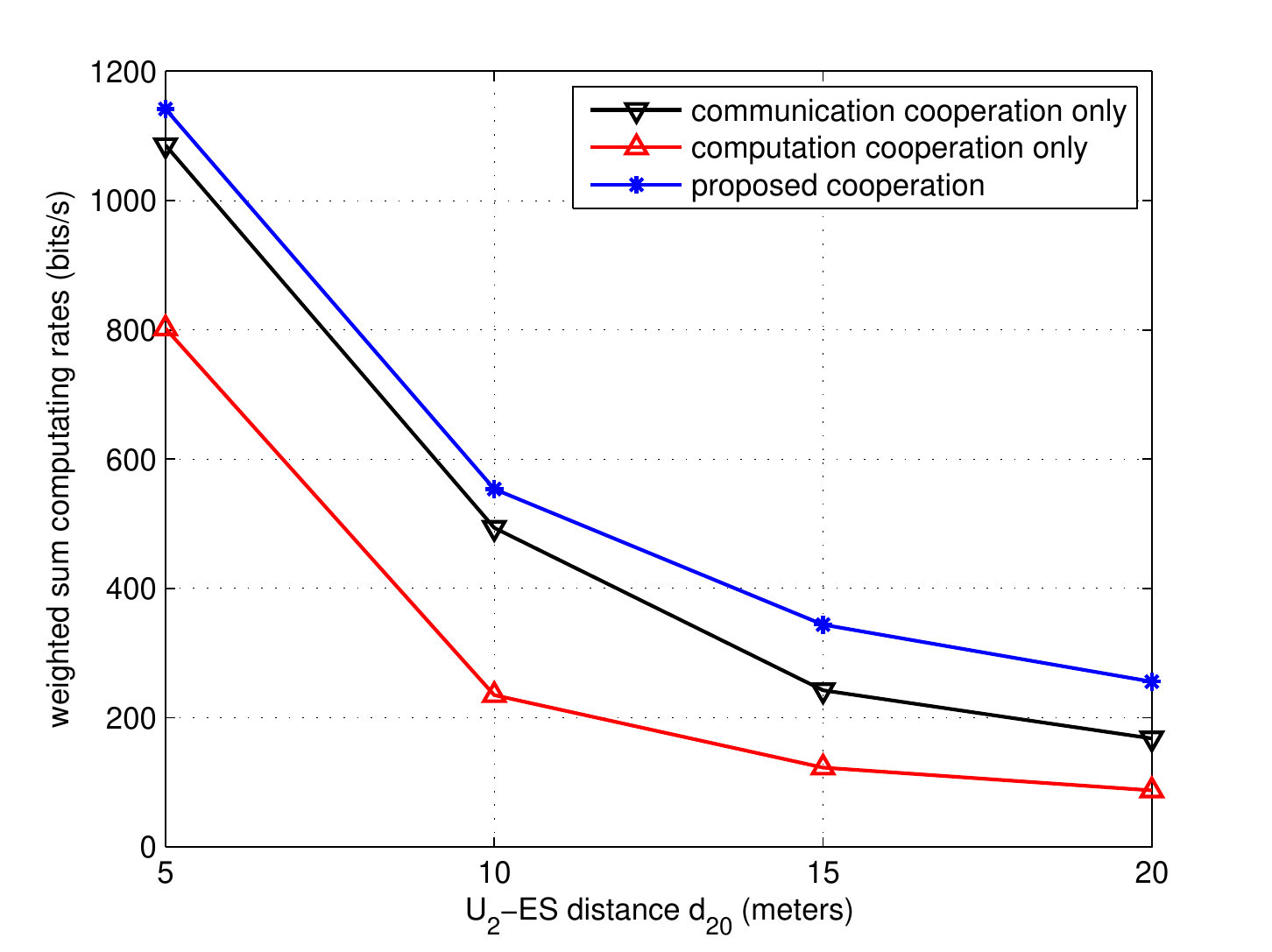}
  \end{center}
  \caption{The impact of $\rm{U_2}$-ES channel ($h_2$) to the optimal WSCR performance.}
  \label{Fig.4}
\end{figure}

We first show the impact of $\rm{U_2}$-to-ES channel to the optimal WSCR performance in Fig.~\ref{Fig.4}. Here, we let ${d_{20}}$ (the distance between $\rm{U_2}$ and ES) vary from $5m$ to $20m$, while the channel $h_2$ degrades with $d_{20}$. As expected, the WSCR of all the three schemes degrade with the increase of ${d_{20}}$ due to the larger time and energy consumption required to offload the tasks to the ES. We see that the proposed method evidently outperforms the two benchmark methods, where on average it achieves $27.9\%$ and $137.8\%$ higher WSCR than benchmark 1 and 2, respectively. In particular, benchmark 2 has the worst performance as $\rm{U_1}$ cannot utilize the powerful edge computation. Benchmark 1 has comparable performance with the proposed method but the performance gain becomes larger as the $\rm{U_2}$-ES channel weakens, due to the high cost on task offloading becomes the performance bottleneck of the system.

\vspace{0.5cm}

\begin{figure}
  \centering
  \begin{center}
    \includegraphics[width=0.45\textwidth]{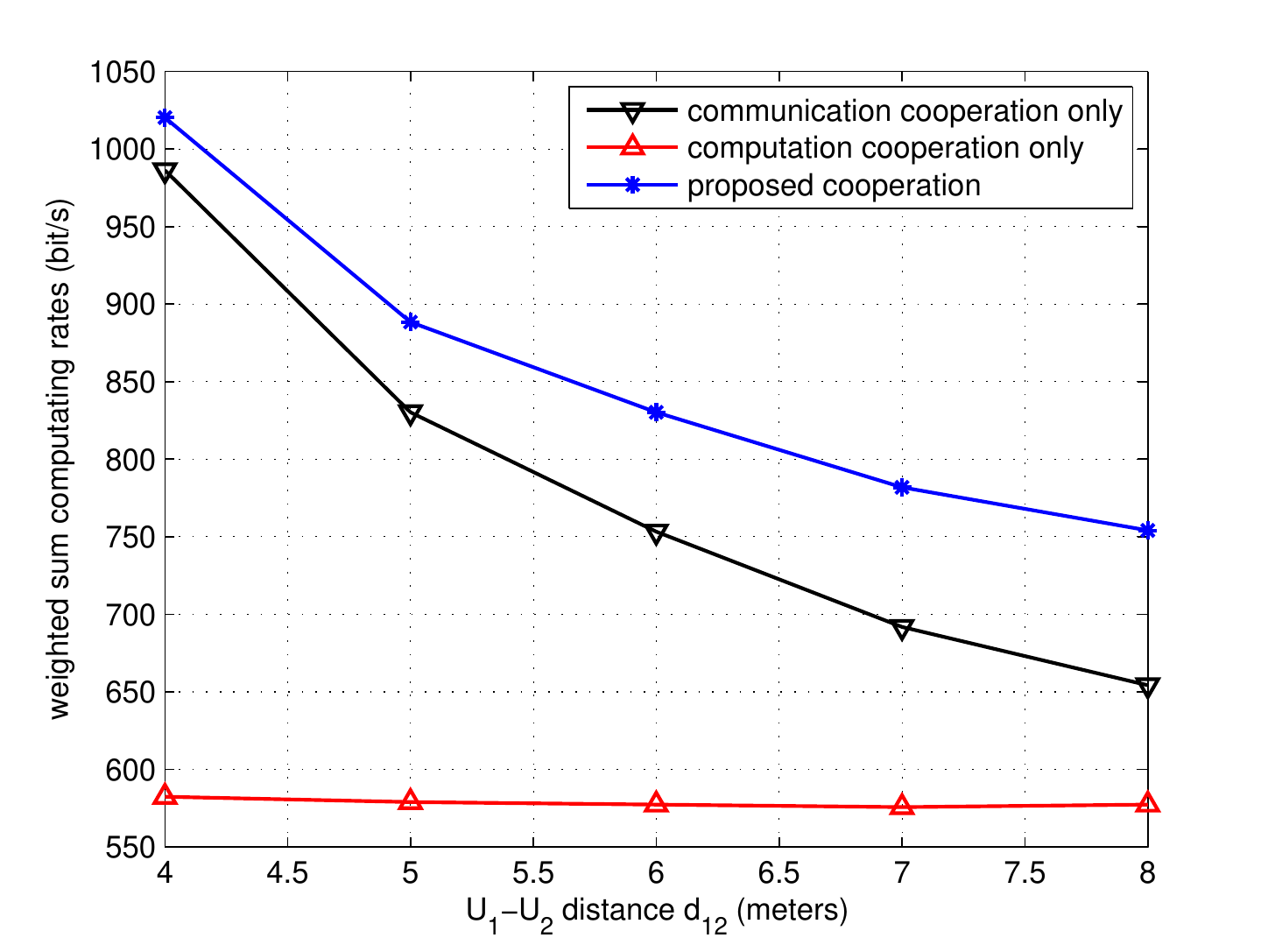}
  \end{center}
  \caption{The impact of inter-user channel ($h_1$) to the optimal WSCR performance. }
  \label{Fig.5}
\end{figure}

In Fig.~\ref{Fig.5}, we further investigate the impact of inter-user channel $h_1$ to the system performance. Here, we set $d_{E,2} = 3m$, $d_{20} = 8m$, and let ${d_{12}}$ increase from $4m$ to $8m$. As expected, with the increase of $d_{12}$, the performances of the proposed method and benchmark 1 decline due to the larger cost on data offloading. Benchmark 2 has relatively steady performance with the change of inter-user channel, because most of the tasks are computed locally for $\rm{U_1}$ at optimum. On average, the proposed cooperation achieves $9.8\%$ and $47.8\%$ higher WSCR than the two benchmark methods.

\vspace{0.5cm}

\begin{figure}
  \centering
  \begin{center}
    \includegraphics[width=0.45\textwidth]{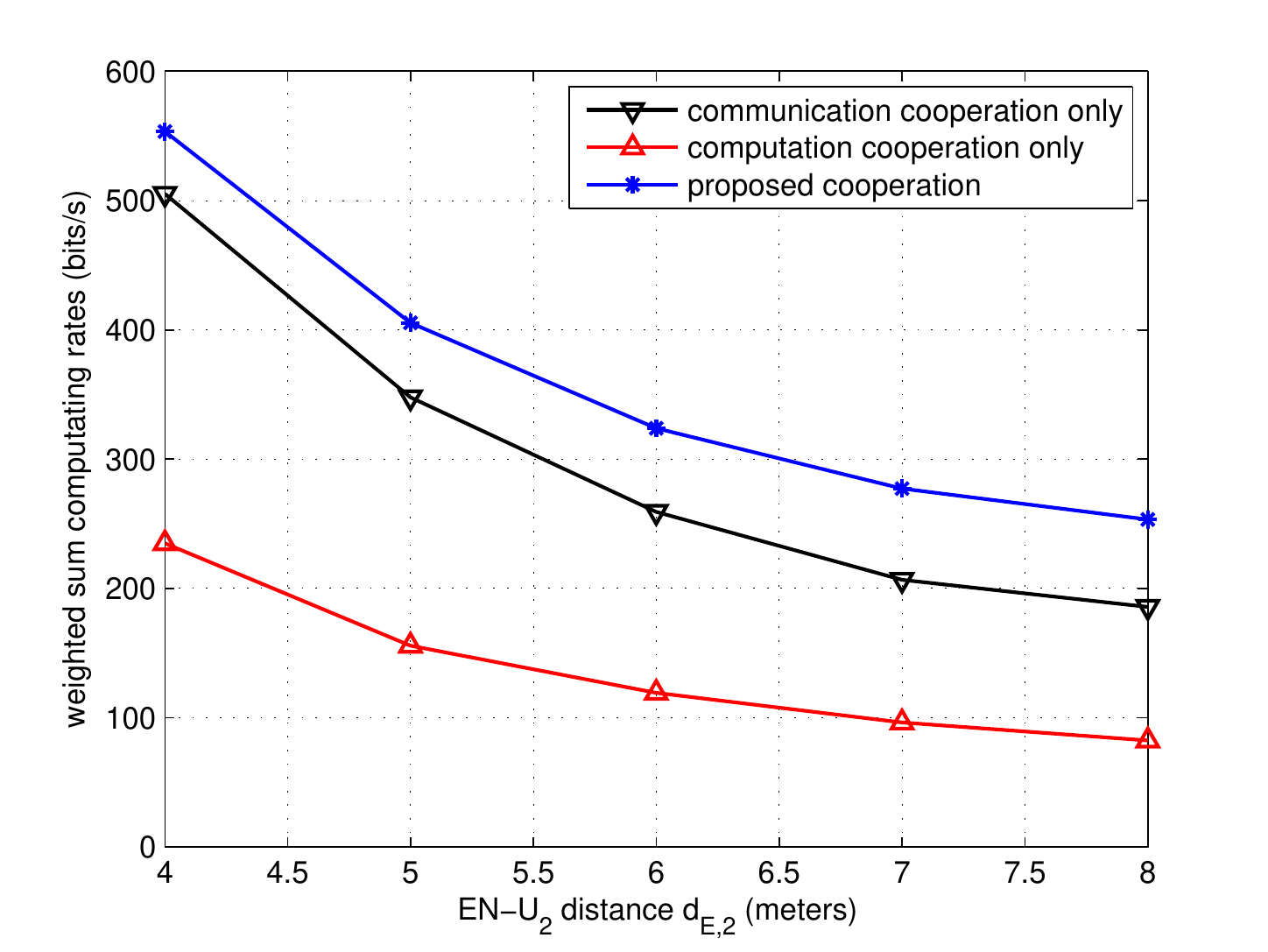}
  \end{center}
  \caption{The impact of EN-U2 channel to the optimal WSCR performance. }
  \label{Fig.6}
\end{figure}

Fig.~\ref{Fig.6} shows the impact of EN-$\rm{U_2}$ channel ${d_{E,2}}$ to the system performance. Here, we vary ${d_{E,2}}$ from $4m$ to $8m$ while the others are default values. Still, the performance of all the methods degrades due to the smaller available energy received by $\rm{U_2}$ under a larger $d_{E,2}$. Nonetheless, on average the proposed scheme still outperforms the two benchmarks by $24.3\%$ and $172.9\%$, respectively.

\vspace{0.5cm}

\begin{figure}
  \centering
  \begin{center}
    \includegraphics[width=0.45\textwidth]{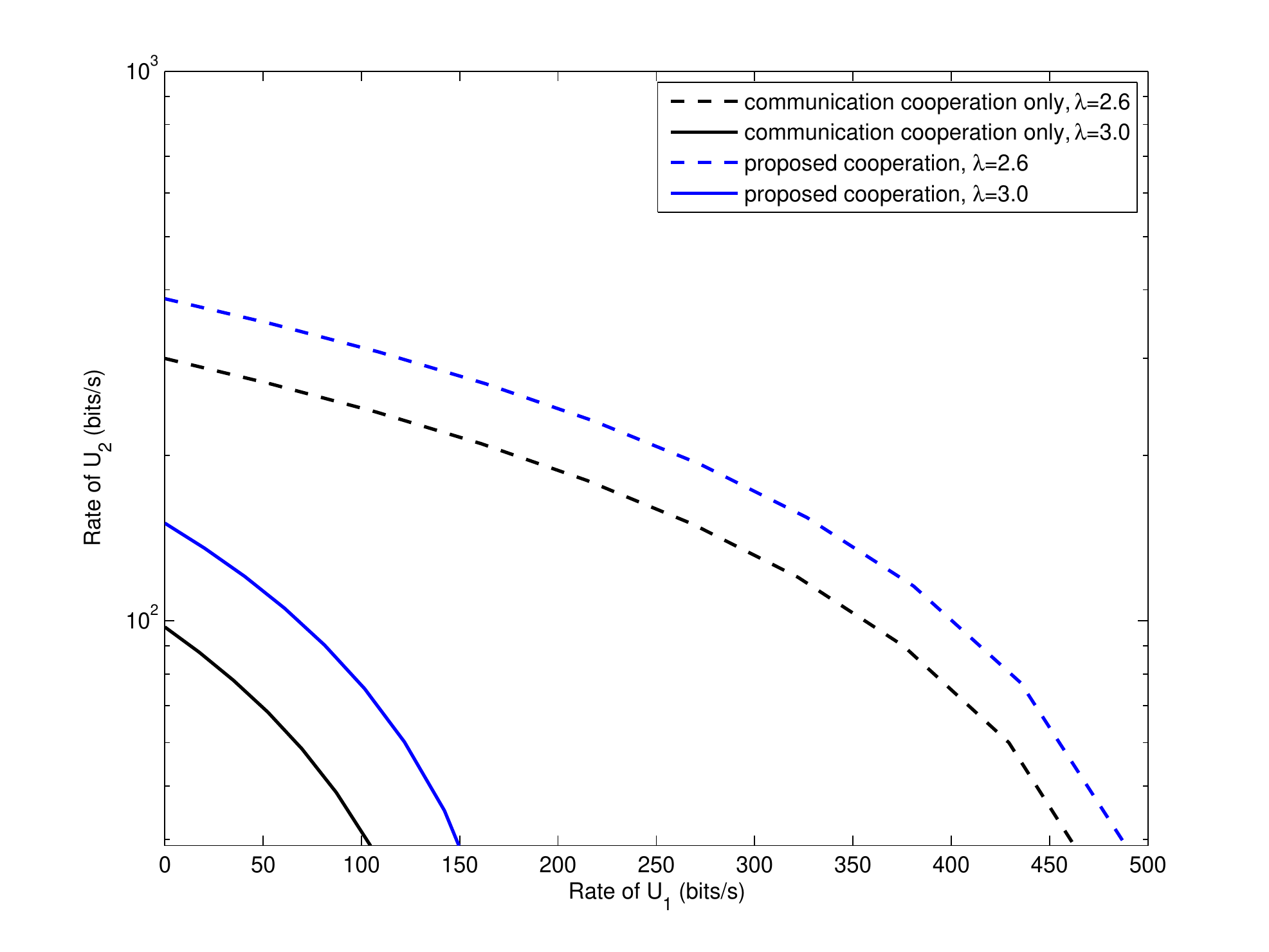}
  \end{center}
  \caption{Rate region comparisons under different path loss exponent.}
  \label{Fig.7}
\end{figure}

Finally, in Fig.~\ref{Fig.7}, we plot computation rate regions (by varying $w_1$ from $0$ to $1$) achieved by the proposed method and the better performing benchmark method, i.e., benchmark 1, under different path-loss factor $\lambda$. Evidently, with a larger $\lambda$, both schemes achieve a smaller rate region due to the smaller energy harvested and larger cost on data offloading. For both values of $\lambda$, the proposed scheme has an evident advantage over benchmark 1, where the regions of the benchmark are inside those of the proposed scheme. This indicates that both users can benefit from the considered cooperation.

The above results verify that the proposed cooperation method has strong flexibility in adapting the resource allocation to the change of network parameters for supporting high-performance computation service. In particular, not only the weak user $\rm{U_1}$, but also the helper can benefit from the proposed joint communication and computation cooperation. The performance advantage is especially evident when the offloading channel, either due to $\rm{U_1}$-$\rm{U_2}$ or $\rm{U_2}$-ES, is weak.

\section{Conclusions}
In this paper, we investigated a joint communication and computation cooperation method in a two-user wireless powered MEC system. We formulated an optimization problem to maximize the two users' weighted sum computation rates and proposed an efficient method to solve it optimally. Simulation results show that the proposed cooperation method can effectively enhance the computation performance of the system under different network setups compared to other representative benchmark methods, especially when task offloading is costly for either user. Besides, both users can benefit from the proposed cooperation.

%\end{IEEEbiography}
\end{document}